\documentclass[twocolumn,aps,pra,amsmath,amssymb,superscriptaddress,longbibliography,floatfix]{revtex4-1}
\usepackage[T1]{fontenc}
\usepackage[latin9]{inputenc}
\usepackage{amsmath}
\usepackage{graphicx}

\makeatletter
\usepackage{dcolumn}
\usepackage{bm}
\usepackage{times}
\usepackage{color}

\makeatother

\begin{document}
\title{Orbital Rotations induced by Charges of Polarons
and Defects in Doped Vanadates}
\author{Peter Horsch}
\affiliation{Max-Planck-Institut für Festkörperforschung, Heisenbergstrasse 1,
D-70569 Stuttgart, Germany }
\author{Andrzej M. Ole\'{s}}
\affiliation{Max-Planck-Institut für Festkörperforschung, Heisenbergstrasse 1,
D-70569 Stuttgart, Germany }
\affiliation{%
\mbox{%
Institute of Theoretical Physics, Jagiellonian University, Profesora
Stanis\l awa \L ojasiewicza 11, PL-30348 Kraków, Poland%
}}
\author{Adolfo Avella}
\affiliation{Dipartimento di Fisica ``E.R. Caianiello'', Università degli Studi
di Salerno, I-84084 Fisciano (SA), Italy}
\affiliation{CNR-SPIN, UoS di Salerno, I-84084 Fisciano (SA), Italy}
\affiliation{Unità CNISM di Salerno, Università degli Studi di Salerno, I-84084
Fisciano (SA), Italy}
\date{\today}
\begin{abstract}

We explore the competiton of doped holes and defects that leads to the
loss of orbital order in vanadate perovskites. In compounds such as
La$_{1-{\sf x}}$Ca$_{\,\sf x}$VO$_3$ spin and orbital order result
from super-exchange interactions described by an extended
three-orbital degenerate Hubbard-Hund model for the vanadium $t_{2g}$
electrons. Long-range Coulomb potentials of charged Ca$^{2+}$ defects
and $e$-$e$ interactions control the emergence of defect states inside
the Mott gap. The quadrupolar components of the Coulomb fields of doped
holes induce anisotropic orbital rotations of degenerate orbitals.
These  rotations  modify the spin-orbital polaron
clouds and compete with orbital rotations induced by defects.
Both mechanisms lead to a mixing of orbitals,
and cause the suppression of the asymmetry of
kinetic energy in the $C$-type magnetic phase.
We find that  the gradual decline of orbital order with doping,
a characteristic feature of the vanadates, however, has its origin not
predominantly in the charge carriers, but in the off-diagonal couplings
of orbital rotations induced by the charges of the doped ions.
\end{abstract}
\maketitle

\section{Introduction}
\label{sec:intro}

The discovery that doping holes (or electrons) into Mott insulators
(MIs), formed by CuO$_{2}$ layers, not just leads to a metallic state
and to the decay of the antiferromagnetic (AF) order but also
yields high temperature superconductivity \citep{Mul86} was a great
surprise. Very early, it was proposed that the mechanism of high
temperature superconductivity originates from the strong electron
correlations, intrinsic to the MI \mbox{\citep{And87,Fra15,Kei15},}
rather than from the exchange of phonons. The discovery triggered a
systematic study of transition metal oxides \citep{Kho14,Ima98}, many
of them MIs, also to achieve deeper insights into the fundamental open
questions related to doped MIs. More recently, time dependent phenomena
came into focus with the discovery of ultracold
fermion systems in optical lattices \mbox{\citep{Tro08,Gru18,Gru19}}
that represent alternative platforms to study Mott-Hubbard physics.

Spin and/or orbital ordered phases are typical features of
orbitally-degenerate MIs. The motion of doped holes in such compounds
leads to strings of misplaced spins
\citep{Bri70,Liu92,Hor94,Lee06,Oga08,Mou07,spins,Chi19,Bie19} or
misoriented orbitals \citep{Dag08,Wro10,Wro12} or both
\citep{Ish05,Woh09,Ave18,BBO19}. In cuprates, the perturbation induced
by doping holes can be efficiently described in terms of spin-polarons
\citep{Kan89,Mar91} moving in a two-dimensional (2D) spin-1/2 quantum
antiferromagnet. Due to the strong quantum fluctuations, their kinetic
energy is only weakly reduced such that their binding to charged
defects is small. As a consequence, the metal-insulator transition and
the breakdown of AF long-range order occur already at quite low doping.
In contrast, the perovskite vanadates, that reveal strong quantum
orbital fluctuations in certain regimes \citep{Yan04,And07}, remain
insulating up to moderate or even high doping concentrations.
For instance, Y$_{1-{\sf x}}$Ca$_{\sf x}$VO$_{3}$ enters a poor
metallic state only at ${\sf x}\approx0.5$ \citep{Pen99}.
In the parent compounds $R$VO$_{3}$, with $R\!\equiv$ La, Pr, ...,
Tb, Y, and Lu, the $d^{2}$ configuration of V$^{3+}$ ions has
an orbital degeneracy of the $t_{2g}$ electrons and Hund's exchange
stabilizes high spin $S\!=\!1$ states \citep{Miy06}.

\begin{figure}
\noindent \centering{}\includegraphics[width=1\columnwidth]{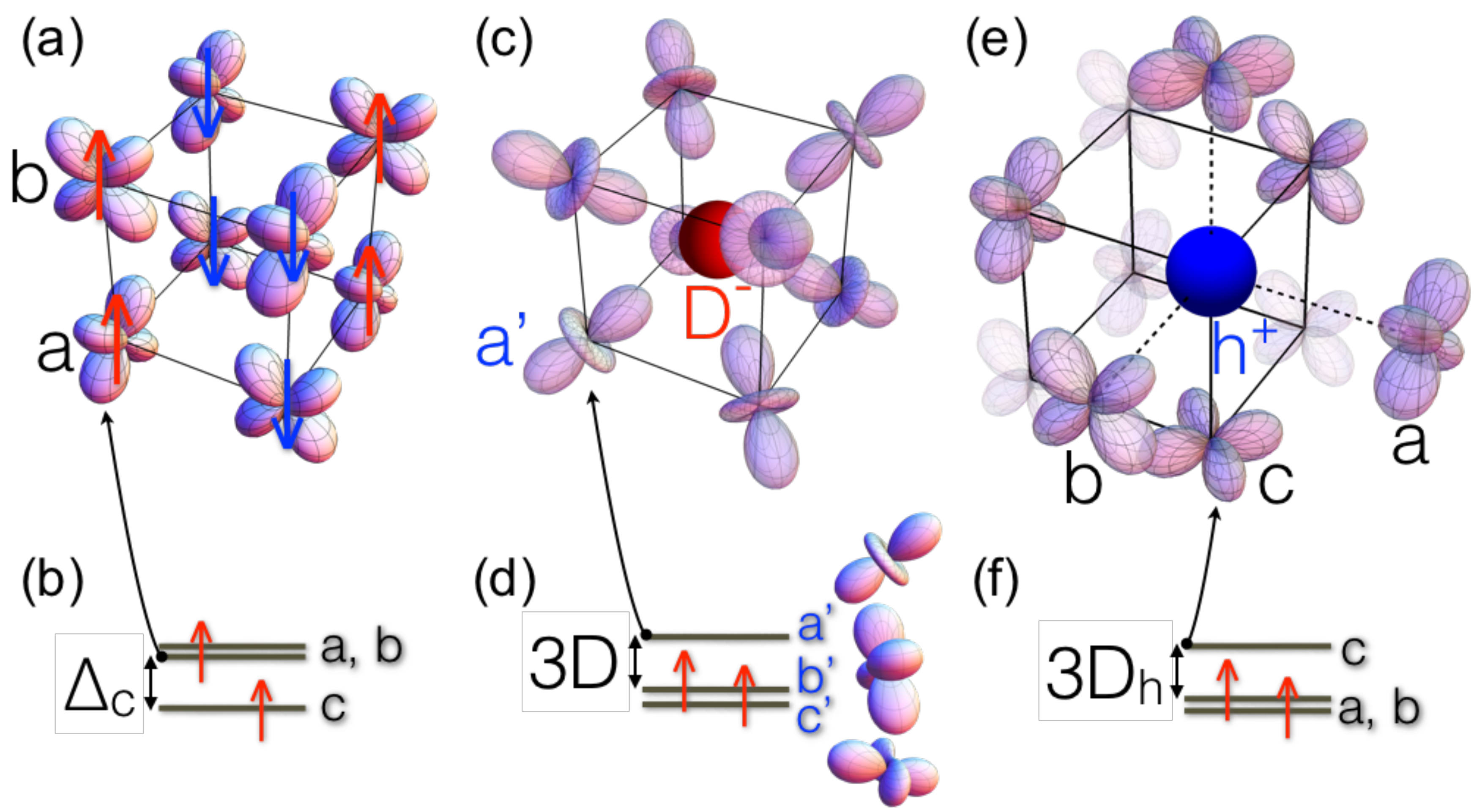}
\caption{
Orbital rotations due to charges of defects and doped holes in
$R$VO$_{3}$:
(a) CS/GO reference state with $C$-type order of $S=1$ spins of V$^{3+}$
($d^{2}$) ions and alternating G-type orbital order of occupied
$a\!\equiv\!yz$ and $b\!\equiv\!xz$ vanadium orbitals. The occupied
$c\!\equiv\!xy$ orbitals (not shown) are stabilized by a crystal field
$\Delta_{c}$, see (b).
Orbital rotations induced by a charged defect (red dot, Ca$^{2+}$
ion) and by the charge of a doped hole (blue dot, V$^{4+}$ ion) on
nearest neighbor V$^{3+}$ ions are shown in (c) and (e), respectively.
For clarity we display only the highest rotated (i.e., unoccupied)
orbital per V$^{3+}$ ion in the strong coupling limit (i.e.,
$\Delta_{c}\ll{\cal D}$ or ${\cal D}_{h}$), whereas (d) and (f)
display the level splittings for moderate values of coupling constants,
${\cal D}$ and ${\cal D}_h$.
\label{fig:1}}
\end{figure}

Orbitally-degenerate Mott insulators display various kinds of spin and
orbital order
\citep{Dag01,Ish97,Fei99,Wei04,Fei97,Ole05,Kha05,Nor08,Zaa09}. $R$VO$_3$
has several phases with the respective transition temperatures
dependent on the radius of the $R$-ions. For instance, all compounds
from La to Tb have a ground state with $C$-type spin and $G$-type
orbital order (CS/GO) \citep{Fuj05}, i.e., with antiferro order in the
$ab$-planes and ferro order along the $c$ axis, respectively, while the
occupation of the topmost occupied $t_{2g}$ orbital alternates between
$a\!=\!yz$ and $b\!=\!xz$ \citep{Kha00} as shown in Fig.~\ref{fig:1}(a).
The same state was found recently in LaVO$_3$ thin films \cite{Lov20}.
Above the Néel temperature $T_{{\rm N}}$, there is a paramagnetic phase
with $G$-type orbital order, which disappears above the orbital order
temperature $T_{{\rm OO}}$. In YVO$_{3}$, and all systems with smaller
$R$ ionic radius, there is a further transition at $T_{\rm CG}$ from the
CS/GO to a complementary GS/CO low-$T$ state \citep{Ren98,Fuj10,Sah17}.
The CS/GO order emerges from the intrinsic super-exchange interactions,
i.e., driven by strong pseudospin 1/2 orbital quantum fluctuations
along the $c$ axis. These orbital fluctuations lead to much stronger FM
spin-couplings in the CS/GO phase \citep{Kha01,Kha04,Ole07}
than expected for  frozen orbitals, i.e., as assumed in
Goodenough-Kanamori rules \citep{Goode}.
YVO$_{3}$ instead has a GS/CO ground state triggered by also present
Jahn-Teller interactions that increase with decreasing $R$-ion radius
\citep{Bla01,Hor03,Ros18,Yan19}. Interestingly,
already traces of Ca-doping switch the GS/CO ground state of YVO$_3$
to the CS/GO-phase \citep{Fuj05,Fuj08,Ree16}, a feature that could be
explained in a model for charged defects that we adopt here
\citep{Hor11,Ave13}.

Here, we explore the stability of the disordered CS/GO-phase and its
gradual decay at large doping. It is the coupling to the extra orbital
degree of freedom in the vanadates that leads to the quenching of the
kinetic energy and to strong localization and binding of polarons by
the Coulomb potential of defects. Yet, this strong localization creates
a new puzzle: \textit{how is then the orbital order destroyed in
these compounds?} As alternative mechanism, the orbital rotations (ORs)
at vanadium ions were identified.
They are induced by the Coulomb fields
of the charged defects \citep{Hor11,Ave13}. It was shown that ORs
are an effective perturbation as each defect is surrounded by eight
nearest vanadium neighbors, see Fig.~\ref{fig:1}(b). It yields a
natural explanation for the gradual suppression of $G$-type orbital
order in vanadates as function of doping \citep{Ave19}, and the absence
of clear signatures of collective phase transitions \citep{Ave15}. In
this paper, we explore a complementary, {\it a priori} equally
important, orbital polarization mechanism triggered by the polaron
charge, see Fig.~\ref{fig:1}(c).
A mechanism of this kind was found essential
by Kilian and Khaliullin \citep{Kil99} in a study of orbital polarons in
the orbital liquid regime of $e_{g}$ orbitals in manganites
\citep{Kim02,Dag04,Gec05} .
For the $t_{2g}$ orbitals of vanadates we find an abrupt
reduction of orbital order caused  by doped holes/polarons beyond
a critical coupling strength. Yet in combination with OR's induced by
the defect charges the gradual decline of orbital order dominates and
is amplified by  OR's due to the polarons.


Despite the cubic structure, the undoped CS/GO state is highly
anisotropic, due to the FM correlations along the $c$-axis, where
strong quantum orbital fluctuations boost the super-exchange
\citep{Yan04,Kha01,Kha04,Ole07}.
Tokura and coworkers \citep{Fuj06} observed that the anisotropy
ratio $A_{opt}$ of optical weights along $z$ and $x$ axis changed
from about two to one at large doping in the CS phase. A goal of
our work is to shed light on this puzzle by studying the asymmetry
in the kinetic energy $A=K_{z}/K_{x}$ that is strictly related to
$A_{opt}$ \citep{Kha04}. The case of the vanadates is puzzling
as the isotropy of kinetic energy is observed at doping concentrations
where the anisotropic magnetic CS order still persists.

The article continues in Section II with a brief description of the minimal model for
the doped vanadate Mott insulators. The main focus here is on the
orbital rotation terms that control the orientation of vanadium orbitals,
and are a consequence of the Coulomb fields of defects and doped holes or electrons.
In Section III the effect of  orbital rotations on the occupation
of orbitals, the magnetic and the orbital order is studied. In
Section IV we present our conclusions. An Appendix contains
further details of the multi-orbital Hubbard-Hund interaction, the
Jahn-Teller and other small terms,
as well as the derivation of the orbital polarization terms.

\section{The multi-orbital model for doped vanadate Mott insulators}
\label{sec:model}

The minimal Hamiltonian that describes the $t_{2g}$ electrons, the
Mott gap, and the defect states in $R_{1-{\sf x}}$Ca$\,_{\sf x}$VO$_{3}$
is \citep{Ave13}:
\begin{equation}
{\cal H}_{t2g}\!={\cal H}_{{\rm Hub}}+\sum_{mi}v(r_{mi})\hat{n}_{i}
+\sum_{i<j}v(r_{ij})\hat{n}_{i}\hat{n}_{j}+{\cal H}_{{\rm pol}}.
\label{Ht2g}
\end{equation}
It includes an extended 3-band Hubbard model ${\cal H}_{{\rm Hub}}$
\citep{Pen97,Dag10} that describes the electronic multiplet structure
of the V$^{3+}$ ions \citep{Hor11} and the different phases of
the parent compounds. For a first orientation the details of this
term can be ignored. They are described, however, in Appendix A.
The 2$^{nd}$ term in Eq.~(\ref{Ht2g}) describes the Coulomb potentials
of D$^{-}$ defects, that have an effective negative charge and represent,
for instance, Ca$^{2+}$ substituting $R^{3+}$ ions. Defects attract
doped holes and strongly repel electrons of V-ions in the vicinity and
shift these states from the lower Hubbard band into the Mott gap
\citep{Ave18}. The 3$^{rd}$ term, the $e$-$e$ interaction, leads to the
screening by $t_{2g}$ electrons and doped holes. Both terms are
determined by the Coulomb field, $v(r)\equiv{e^2}/{\varepsilon_{c}r}$,
where $\varepsilon_{c}\simeq5$ is the dielectric constant of the
core electrons \citep{Hor11} and $r$ is the distance between charges of:
(i)~a defect D$^{-}$ at site $m$ and $t_{2g}$ electrons at a V ion at
site $i$, i.e., %
\mbox{%
$r_{mi}\!=|\mathbf{R}_{m}-\mathbf{r}_{i}|$%
}, and (ii)~two V ions at sites $i$ and $j$ with %
\mbox{%
$r_{ij}\!=|\mathbf{r}_{i}-\mathbf{r}_{j}|$%
}. $\hat{n}_{i}=\sum_{\alpha\sigma}\hat{n}_{i\alpha\sigma}$ is the
$t_{2g}$ electron charge operator with %
\mbox{%
$\hat{n}_{i\alpha\sigma}\!=\!\hat{d}_{i\alpha\sigma}^{\dagger}\hat{d}_{i\alpha\sigma}$%
} and orbital flavors $\alpha\!=\!\{yz,zx,xy\}\!\equiv\!\{a,b,c\}$
\citep{Kha00}, see Fig.~\ref{fig:1}(a).

Central to our discussion are the orbital polarization terms,
${\cal H}_{{\rm pol}}\equiv
\mathcal{H}_{{\rm pol}}^{(1)}+\mathcal{H}_{{\rm pol}}^{(2)}$.
They describe the OR and the redistribution of $t_{2g}$ electronic
charge at V-ions, induced by the Coulomb fields of defects and
doped holes. These terms appear in addition to the monopole
terms that are already contained in the minimal model.
The orbital polarization term $\mathcal{H}_{{\rm pol}}^{(1)}$,
due to the charged defect \citep{Ave13} reads as:
\begin{equation}
\mathcal{H}_{{\rm pol}}^{(1)}={\cal D}\!\!\sum_{{m,i\in\mathcal{C}_{m}\atop \alpha,\beta,\sigma}}\!\lambda_{\alpha\beta}^{\mathbf{d}}\;
\delta_{\mathbf{d},\mathbf{r}_{i}\!-\!\mathbf{R}_{m}}%
\hat{d}_{i\alpha\sigma}^{\dagger}\hat{d}_{i\beta\sigma},
\label{pol}
\end{equation}
where the coupling constant ${\cal D}$ is defined by the matrix element
\mbox{$\langle i\alpha|v(|\mathbf{r}\!-
\!\mathbf{R}_{m}|)|i\beta\rangle\equiv{\cal D}\lambda_{\alpha\beta}^{{\bf d}}$}
in the basis $\alpha=\{a,b,c\}$, see Fig.~\ref{fig:1}(a), with
$\mathbf{d}\!=\!\mathbf{r}_{i}\!-\!\mathbf{R}_{m}$. We shall treat
${\cal D}$ as a free parameter; a typical value is ${\cal D}\approx50$
meV \citep{Ave19}. The effect of orbital rotation is short-ranged
and affects only the 8 V ions of the defect cube $C_{m}$ of the defect
$m$. The matrix elements $\lambda_{\alpha\beta}^{{\bf d}}$ are traceless,
like the 3-flavor SU(3) matrices \citep{Gel64,Mae04}. They depend
on the diagonal axis ${\bf d}$ in the defect cube, i.e.,
\[
\lambda_{\alpha\beta}^{{\bf d}}=\left(\!\begin{array}{ccc}
0 & 1 & 1\\
1 & 0 & 1\\
1 & 1 & 0
\end{array}\!\right),\left(\!\begin{array}{ccc}
0 & 1 & {\bar{1}}\\
1 & 0 & {\bar{1}}\\
{\bar{1}} & {\bar{1}} & 0
\end{array}\!\right),\left(\!\begin{array}{ccc}
0 & {\bar{1}} & 1\\
{\bar{1}} & 0 & {\bar{1}}\\
1 & {\bar{1}} & 0
\end{array}\!\right),\left(\!\begin{array}{ccc}
0 & {\bar{1}} & {\bar{1}}\\
{\bar{1}} & 0 & 1\\
{\bar{1}} & 1 & 0
\end{array}\!\right)
\]
for $\mathbf{d}||(111),(11{\bar{1}}),(1{\bar{1}}1)$ and $({\bar{1}}11)$,
respectively. In the large ${\cal D}$ limit, the unoccupied orbital
$|a'\rangle\!=(|a\rangle\!+|b\rangle\!+|c\rangle)/\sqrt{3}$ shown
in Fig.~\ref{fig:1}(c) follows from $\mathcal{H}_{{\rm pol}}^{(1)}$
alone. It has the largest overlap with the negative defect and thus
the highest energy.

The orbital polarizations $\mathcal{H}_{{\rm pol}}^{(2)}$  induced
within the $t_{2g}$ orbitals of V-ions
by the quadrupolar components of the Coulomb fields  of doped holes
 is the central issue of this article. This
perturbation of the orbital order, as well as the competition with
orbital rotations induced by defects,  has not been explored  before.
The perturbation due to the polaron charge
results from $e$-$e$ interactions, where $n_{0}-\langle\hat{n}_j\rangle$
measures the hole-density relative to the undoped system:
\begin{equation}
\mathcal{H}_{{\rm pol}}^{(2)}={\cal D}_{h}\!\!\sum_{{m,i\in\mathcal{N}_{j}\atop \alpha,\beta,\sigma}}\!\zeta_{\alpha\beta}^{\mathbf{h}}\,\left(
n_{0}-\langle\hat{n}_{j}\rangle\right)\hat{d}_{i\alpha\sigma}^{\dag}\hat{d}_{i\beta\sigma}^{}.
\label{hol}
\end{equation}
Here, $n_{0}=2$ and we restrict the sum over $\mathbf{r}_{i}$ to a
neighborhood $N_{j}$ that includes the 6 V neighbors of the hole at $j$
as shown in Fig.~\ref{fig:1}(e).
A detailed derivation is given in the Appendix.
The coupling constant ${\cal D}_{h}$ is
defined by the matrix element of the field of the hole at $\mathbf{r}_j$:
\mbox{$\langle i\alpha|v(|\mathbf{r}_i-\mathbf{r}_j|)|i\beta\rangle
-v(|\mathbf{h}|)\delta_{\alpha\beta}\equiv{{\cal D}_{h}}\zeta_{\alpha\beta}^{{\bf h}}$},
with respect to the orbital basis $a=yz$, $b=zx$, and $c=xy$:
\[
\zeta_{\alpha\beta}^{{\bf h}}=\left(\begin{array}{ccc}
2 & 0 & 0\\
0 & {\bar{1}} & 0\\
0 & 0 & {\bar{1}}
\end{array}\right),\left(\begin{array}{ccc}
{\bar{1}} & 0 & 0\\
0 & 2 & 0\\
0 & 0 & {\bar{1}}
\end{array}\right),\left(\begin{array}{ccc}
{\bar{1}} & 0 & 0\\
0 & {\bar{1}} & 0\\
0 & 0 & 2
\end{array}\right),
\]
where $\zeta_{\alpha\beta}^{{\bf h}}$ depends on the axis $\mathbf{h}||(100),(010)$
and $(001)$, respectively.
That is, for a $(001)$ V-ion neighbor of a hole at $\mathbf{r}_{j}$,
the $c$ orbital energy is raised by 2${\cal D}_{h}$, while $a(b)$ are
lowered by ${\cal D}_{h}$, see Fig.~\ref{fig:1}(f). This term
frustrates the polarization around charged defects \eqref{pol}.



\begin{figure}[t!]
\noindent \centering{}\includegraphics[width=1\columnwidth]{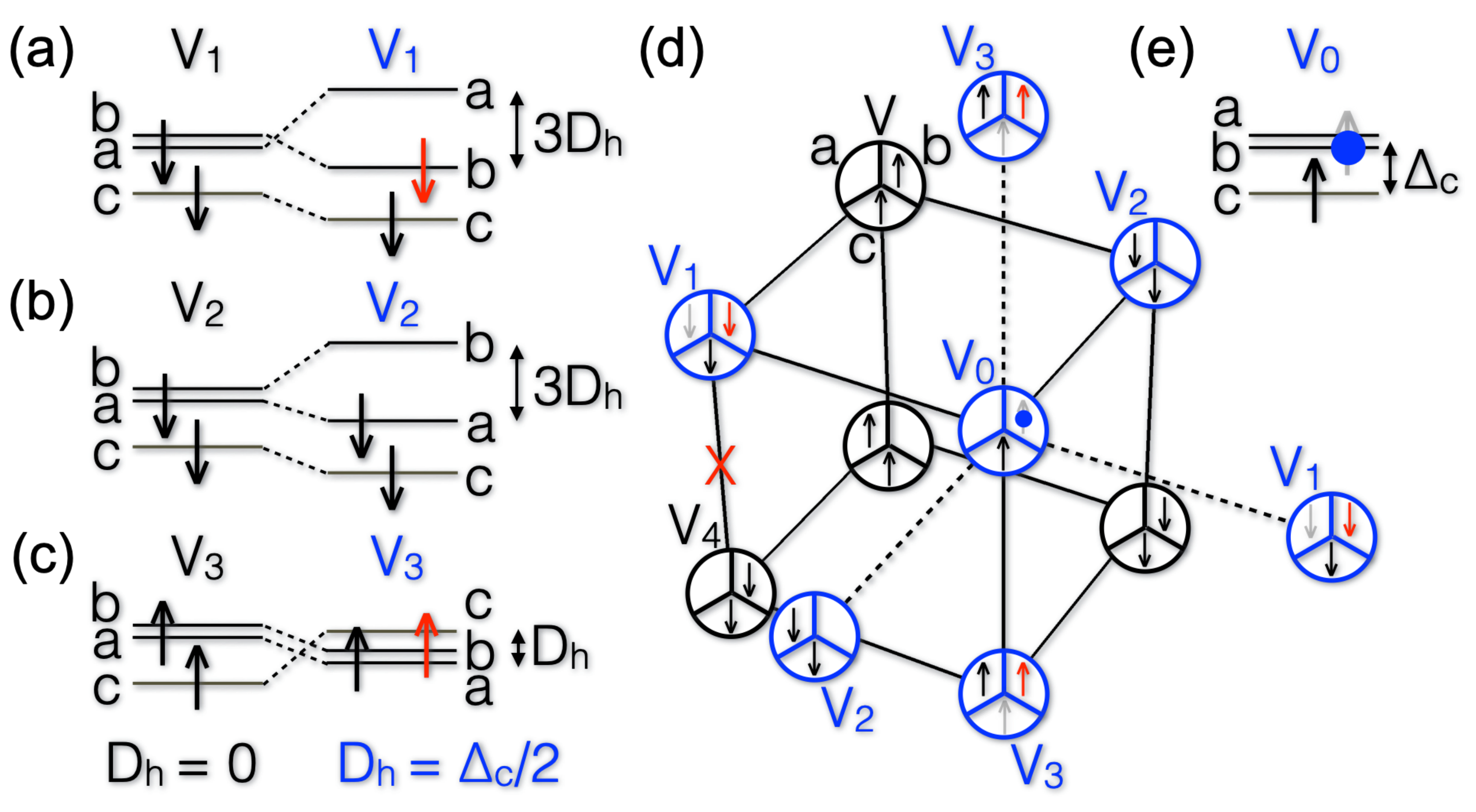}\caption{(a-c) Change of level splitting of neighbor ions V$_{n}$ of a hole
at V$_{0}$ for two different cases: $D_{h}=0$ and large $D_{h}$.
(d) Ions in $x(y,z)$ direction, respectively, from the polaron center
at V$_{0}$ carry labels $n=1(2,3)$ . Arrows indicate the occupation
of orbitals and $a(b,c)$ are orbital labels. A cross indicates the
blocking of superexchange on bond V$_{1}$-V$_{4}$ due to an $a$
to $b$ orbital excitation on ion V$_{1}$ induced by the charge in
the center of the polaron at V$_{0}$ (e) Occupation of the levels
at the V$_{0}$ site, where the hole (blue dot) generating the polarization
cloud resides. Grayed out arrows indicate electrons either removed
by doping (V$_{0}$) or moved to a different orbital (V$_{1}$ and
V$_{3}$). Red arrows indicate the final orbital destination of removed
(grayed out) electrons. \label{fig:1SM}}
\end{figure}

The level splittings induced by the polaron charge are illustrated
in Fig.~\ref{fig:1SM} where the hole was inserted into the $b$
orbital of the ion V$_{0}$ in the  center of the polaron.
In the GO state, all nearest
neighbors of V$_{0}$ have the $c$ and $a$ orbital occupied. The
actions of $\mathcal{H}_{{\rm pol}}^{(2)}$ along the different
cubic directions are different, however. To see
this  anisotropy of the orbital polaron it is useful to consider the in matrix notation
of $\zeta_{\alpha\beta}^{{\bf h}}$ which depends on the axis $\mathbf{h}||(100),(010)$
and $(001)$.
For instance, at the $(001)$ vanadium-ion neighbor,
labeled $V_{3}$, of the doped hole at the ion $V_{0}$ the $c$ orbital energy
is raised by 2${\cal D}_{h}$, while $a(b)$ are lowered by ${\cal D}_{h}$.
This term eliminates the $a/b$-orbital polarization at $V_{3}$ ions.
The other cubic directions are different, for instance along the $b$
axis the occupation of V$_{2}$ does not change at all. Yet along
the $a$-axis the $a$ orbital of V$_{1}$ is raised by $2D_{h}$
while the others are lowered. This leads to switching from $a$ to
$b$ occupation as displayed in V$_{1}$. We note that this switching
is particularly harmful for the $a/b$ orbital order, as it favors
the inverted order.

An important consequence of the orbital excitation from $a$ to $b$
occupation of ion V$_1$ when ${\cal D}_h$ crosses the switching value is
a blocking of orbital super-exchange processes along the $z$ direction.
This is indicated by a red cross on the V$_{1}$-V$_{4}$ bond. Since
the same occurs on the complementary $z$-bonds centered at two equivalent
V$_{1}$ ions the total energy increase, or loss of negative super-exchange
energy, corresponds to 4 broken bonds. Each bond corresponding to
a virtual kinetic energy of $\Delta_{0}=E_{kin}^{0}=0.10$ eV for
$t=0.2$ eV. Thus the total increase of kinetic energy is then 0.4
eV, which coincides with the kinetic energy change per polaron ,
as we shall see further below.

Moreover, the occupation change at V$_{1}$ does not involve the
change of crystal field energies, although there are some changes
of Jahn-Teller energies which we neglect here in the discussion
(but not in the calculations). The latter are however small compared to
the change of super-exchange energy. Therefore, we can conclude that
the crossover scale ${\cal D}_h^c$ is not determined by crystal field
and/or Jahn-Teller terms, but by the interplay of the OR terms and
the super-exchange energy, that is, the scale is determined by the
quantum dynamics of the electrons, and is confirmed by our
numerical study.


\section{Orbital rotations induced by defects and polarons: Results}
\label{sec:results}

In this study, we explore the spin-orbital order of vanadates in the
insulating regime, where doped holes are bound to random defects and
typically prefer a V site on a defect cube; --- which site, depends on
the Coulomb interactions with all other defects and doped holes.
The latter generate polarons and form defect states inside the Mott gap
that persist up to high doping \citep{Ave18}. We calculate the
disordered electronic structure using a variant of the unrestricted
Hartree-Fock method \citep{Miz95,Miz99} that obeys rotational invariance
in both spin and orbital space \citep{Ave13,Ant14}, emphasized as
well in ${\cal H}_{{\rm Hub}}$ \citep{Ole83} and in slave-boson theories
\citep{Ray92,Lee19,Rie19}; this formulation preserves the multiplet
structure of atoms and ions and thereby avoids the shortcomings of the
non-rotational invariant formalism.
\begin{figure}[t!]
\noindent \centering{}\includegraphics[width=0.5\columnwidth]{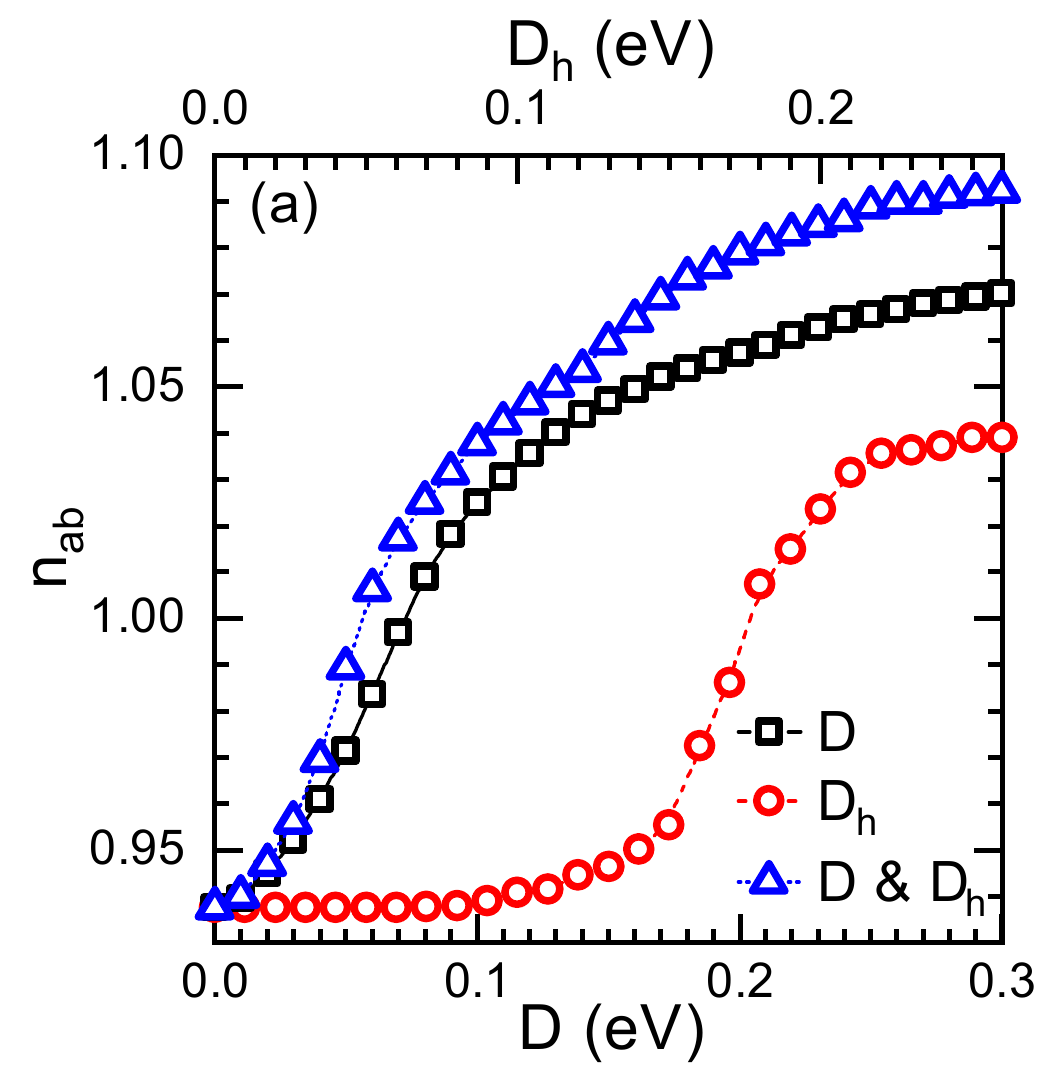}\includegraphics[width=0.5\columnwidth]{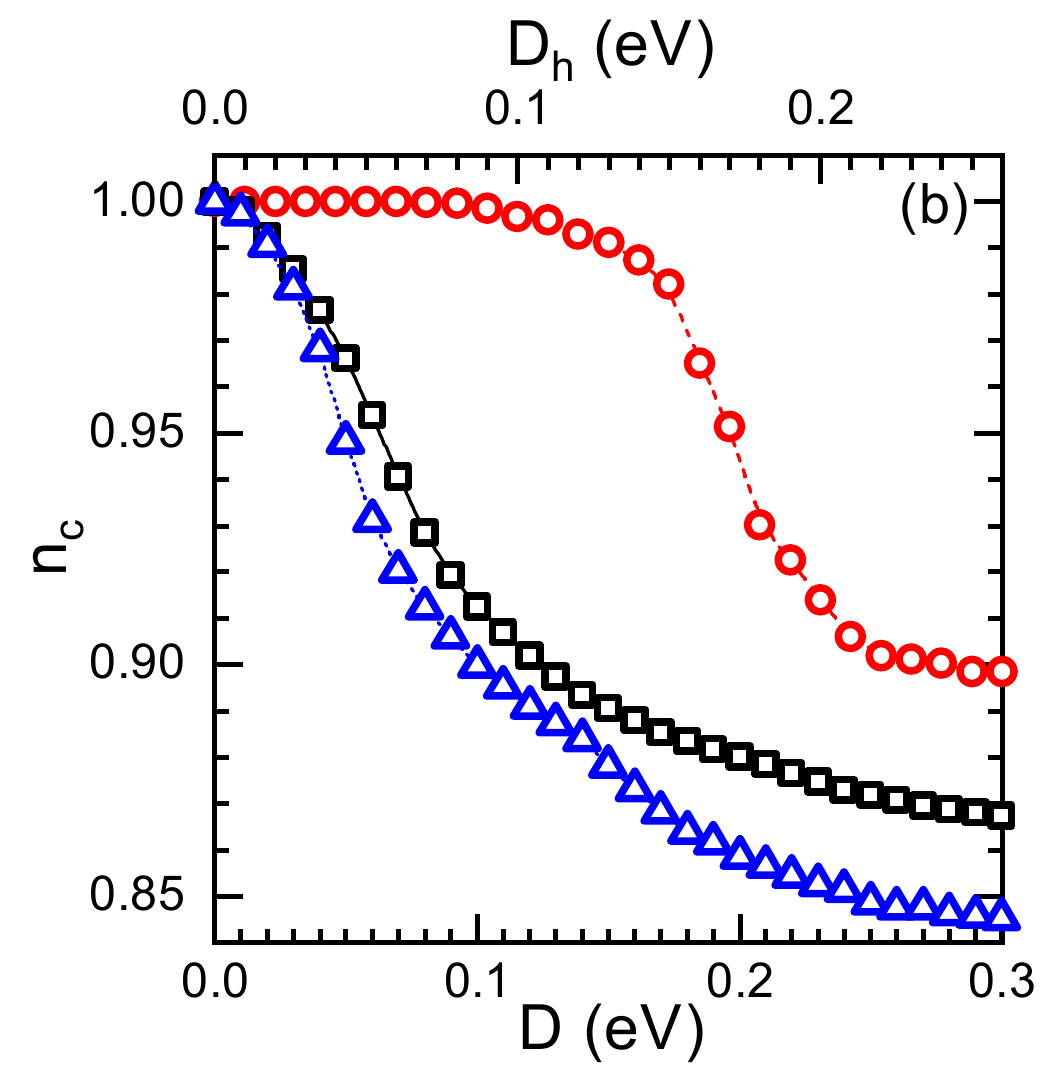}
\caption{Electron densities: (a) $n_{ab}$ for $a$ and $b$ orbitals and (b)
$n_{c}$ for $c$ orbitals, as function of coupling strengths ${\cal D}$
and ${\cal D}_{h}$, respectively, for ${\sf x}=0.0625$. The electron
transfer is compared for three cases, namely the orbital polarization
clouds induced either by defect charges or the polaron charges, as
well as the combined effect. \label{fig:2}}
\end{figure}

We first discuss the effect of $\mathcal{H}_{{\rm pol}}^{(1)}$ and
$\mathcal{H}_{{\rm pol}}^{(2)}$ on the orbital densities, $n_{ab}$
and $n_{c}$. In Fig.~\ref{fig:2}, we monitor three cases, namely
the separate effects of defect and polaron induced OR, as well as
the combined effect, where we use the geometrical relation,
${\cal D}_{h}/{\cal D}=\xi\sim0.87$, defined by the ratio %
\mbox{%
$\xi=d/a$%
} of nearest neighbor V-D ($d$) and V-V ($a$) distances.
For ${\cal D}=0={\cal D}_{h}$, doped holes go
into $a$ and $b$ orbitals due to the crystal field $\Delta_{c}$,
i.e., $n_{ab}=1-{\sf x}$ and $n_{c}=1$. The increase of $n_{ab}$
versus ${\cal D}$ (at ${\cal D}_{h}=0$) can be qualitatively understood
from the rotation of occupied states $\{|c\rangle,|b\rangle\}$ into
$|c'\rangle=(2|c\rangle-|a\rangle-|b\rangle)/\sqrt{6}$ and
$|b'\rangle=(|b\rangle-|a\rangle)/\sqrt{2}$
in the large ${\cal D}$ limit and at $t=0$, see Fig.~\ref{fig:1}(c).

The transfer of holes into the $c$ orbitals due to ${\cal D}_{h}$
(${\cal D}\!=\!0$) is induced by an upward shift of a $c$-orbital
as shown in Fig.~\ref{fig:1}(f). From $\mathcal{H}_{{\rm pol}}^{(2)}$,
one recognizes that, along a 2$^{nd}$ polaron axis, there is no change
of occupation and, along the 3$^{rd}$ axis, there is an interchange
of $a$ and $b$ orbital occupation. It is this latter mechanism
that is particularly harmful for the $a/b$ orbital order.

\begin{figure}[t!]
\noindent \centering{}\includegraphics[width=0.5\columnwidth]{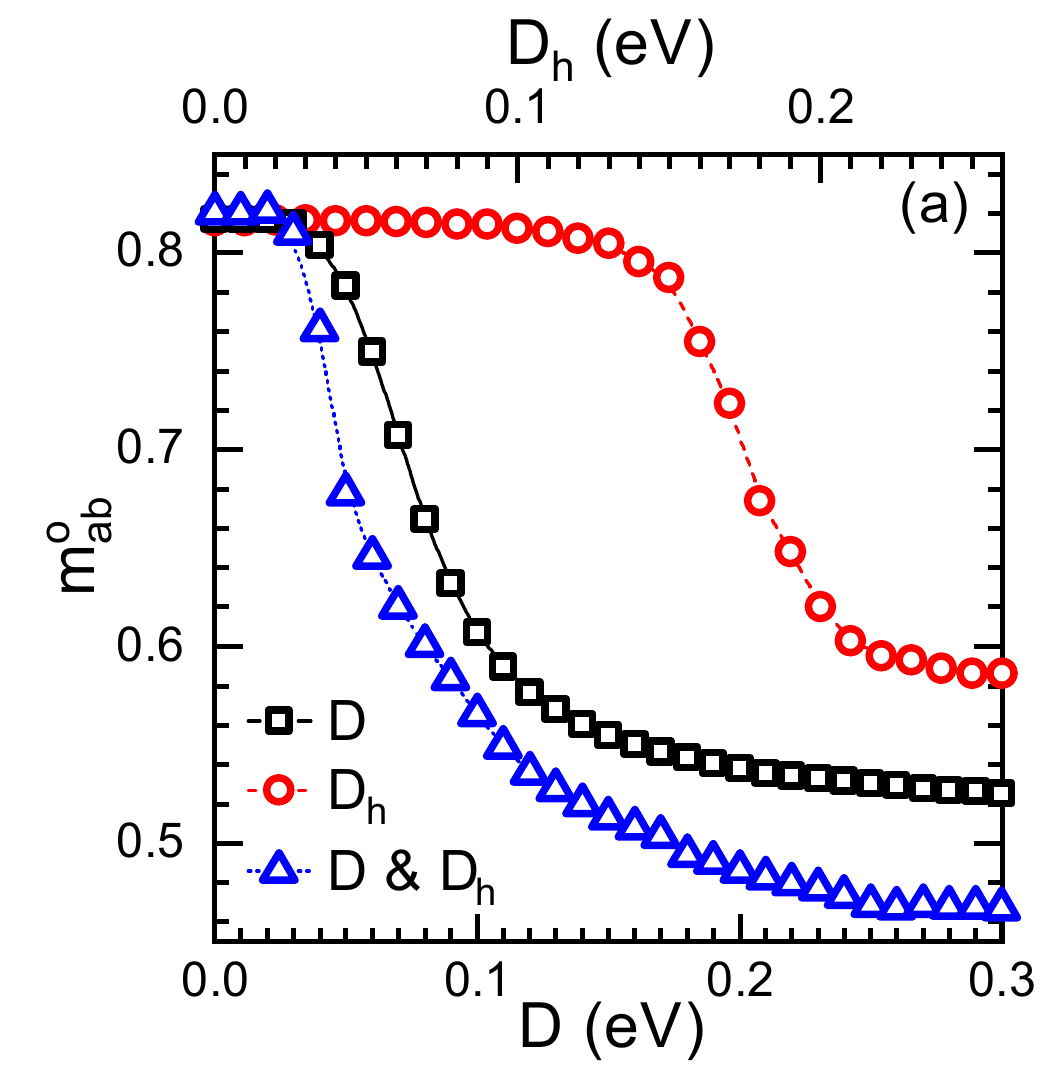}\includegraphics[width=0.5\columnwidth]{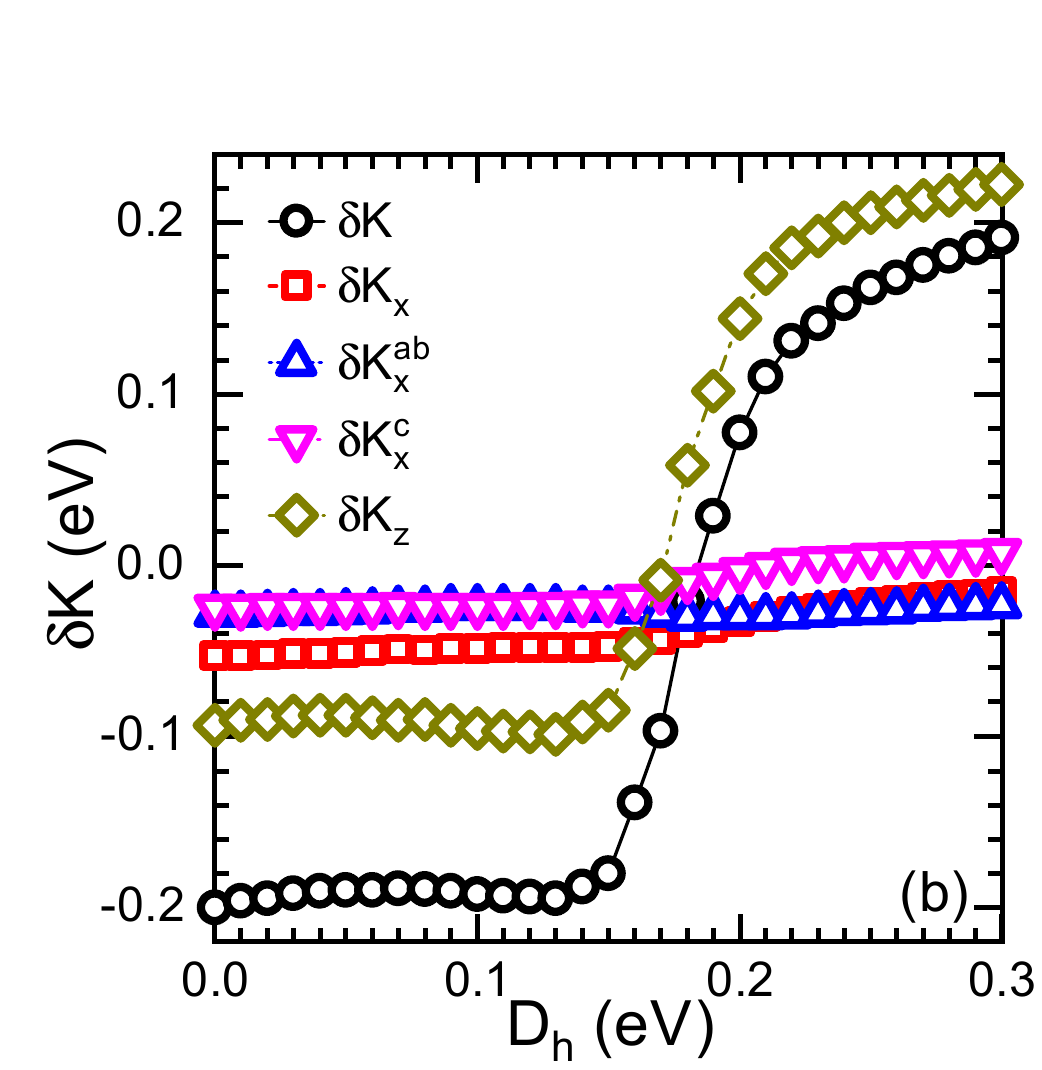}
\caption{Changes of the ground state at doping ${\sf x}=0.0625$ due to orbital
polarization: (a) collapse of G-type orbital order $m_{ab}^{o}$ (4)
for three cases as in Fig.~\ref{fig:2}; (b) kinetic energy change
per doped hole, %
\mbox{%
$\delta K=\delta K_{z}+2\delta K_{x}$%
} versus ${\cal D}_{h}$ for ${\cal D}=0$ with contributions from
hopping along $z$ and $x(y)$, as well as contributions from different
orbitals to $\delta K_{x}=\delta K_{x}^{ab}+\delta K_{x}^{c}$. \label{fig:3}}
\end{figure}

The change of $G$-type orbital order parameter with ${\cal D}$ and
${\cal D}_{h}$ is shown in Fig.~\ref{fig:3}(a). It is determined
by the spatial modulation of the local occupation numbers, $n_{ia}$
and $n_{ib}$,
\begin{equation}
m_{ab}^{o}\equiv\frac{1}{M}\sum_{s=1}^{M}\frac{1}{N}\sum_{i}\left\langle
\hat{n}_{ia}-\hat{n}_{ib}\right\rangle _{s}e^{i\mathbf{Q}_{G}\cdot\mathbf{r}_{i}},
\label{oop}
\end{equation}
where the disorder average is typically taken over $M=100$ defect
realizations $s$ and $\mathbf{Q}_{G}\!\equiv\!(\pi,\pi,\pi)$. Typical
system sizes are $N=4^{3}$ or $8^{3}$. For small ${\cal D}_{h}=\xi{\cal D}$,
the behavior of $m_{ab}^{o}$ is very similar to the pure ${\cal D}$
(${\cal D}_{h}=0$) case. The very different behaviors of the pure
polaron ${\cal D}_h$ (${\cal D}=0$) and defect induced rotation results
from the diagonal versus off-diagonal nature of
$\zeta_{\alpha\beta}^{{\bf h}}$ and $\lambda_{\alpha\beta}^{{\bf d}}$,
respectively. This explains that, in spite of frustration, the
qualitative decay of orbital order of the combined action of
${\cal D}_{h}=\xi{\cal D}$ is similar to the ORs due to defects, yet
the OR due to polarons leads to evident effects for large ${\cal D}$.

The drop of orbital order in Fig.~\ref{fig:3}(a), triggered by the polaron
induced OR, occurs at a crossover scale ${\cal D}_{h}\!\sim0.17$ eV for
${\cal D}\!=\!0$. The crossover scale of $\mathcal{H}_{{\rm pol}}^{(2)}$
is determined by the change of kinetic energy per polaron,
$\delta K\!=\![K({\sf x})-K(0)]/N_{\sf x}$,
where $N_{\sf x}\!=\!{\sf x}N$, see Fig.~\ref{fig:3}(b). The change of
$\delta K\sim0.4$ eV corresponds to the breaking of four super-exchange
bonds along the $z$-direction expected from a switch of $a$ to $b$
occupation at two V-neighbors of a "$b$"-hole.
That is, the stiffness of the magnetic correlations induced by
super-exchange sets the scale, and not the crystal field or the
Jahn-Teller~terms.

\begin{figure}[b!]
\noindent \centering{}\includegraphics[width=0.5\columnwidth]{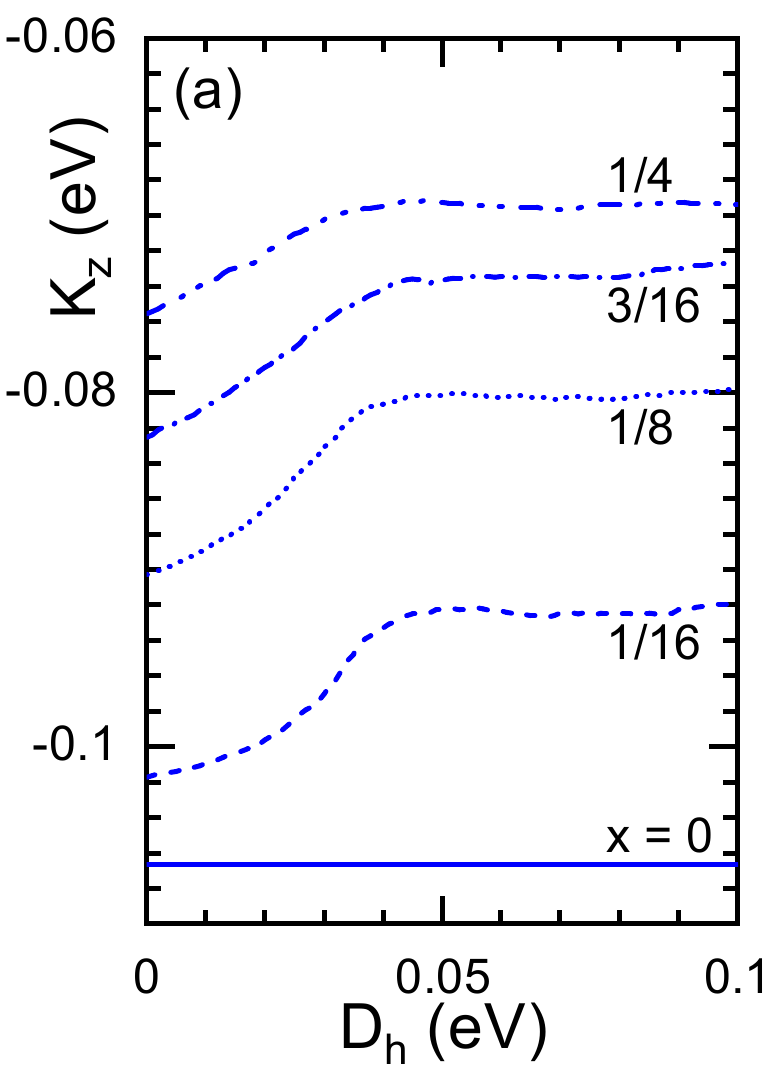}\includegraphics[width=0.5\columnwidth]{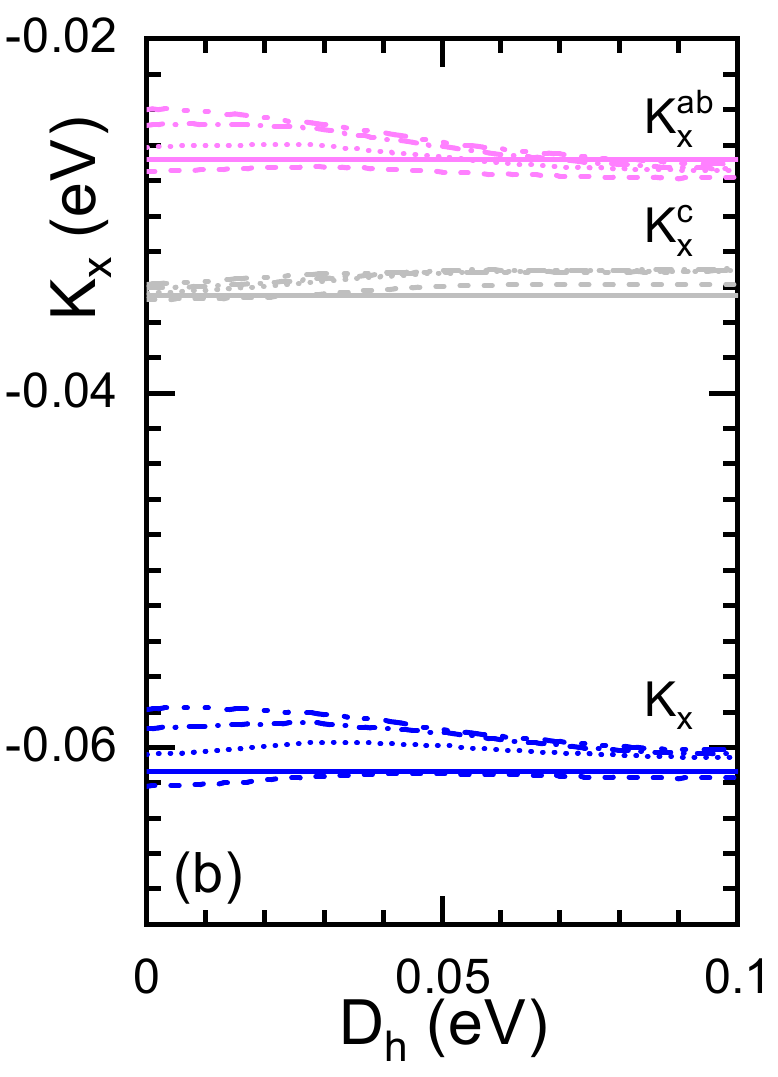}
\caption{(a) Kinetic energy $K_{z}$ versus polarization strength ${\cal D}_h$
(for fixed ${\cal D}=0.05$ eV) and doping ${\sf x}=0,1/16,1/8,3/16$
and $1/4$; and (b) kinetic energy $K_{x}$ along $x$ direction with
its contributions ${K_{x}}^{ab}$ and ${K_{x}}^{c}$ from $a,b$ and
$c$ electrons, respectively. \label{fig:4}}
\end{figure}

\begin{figure}[t!]
\includegraphics[width=\columnwidth]{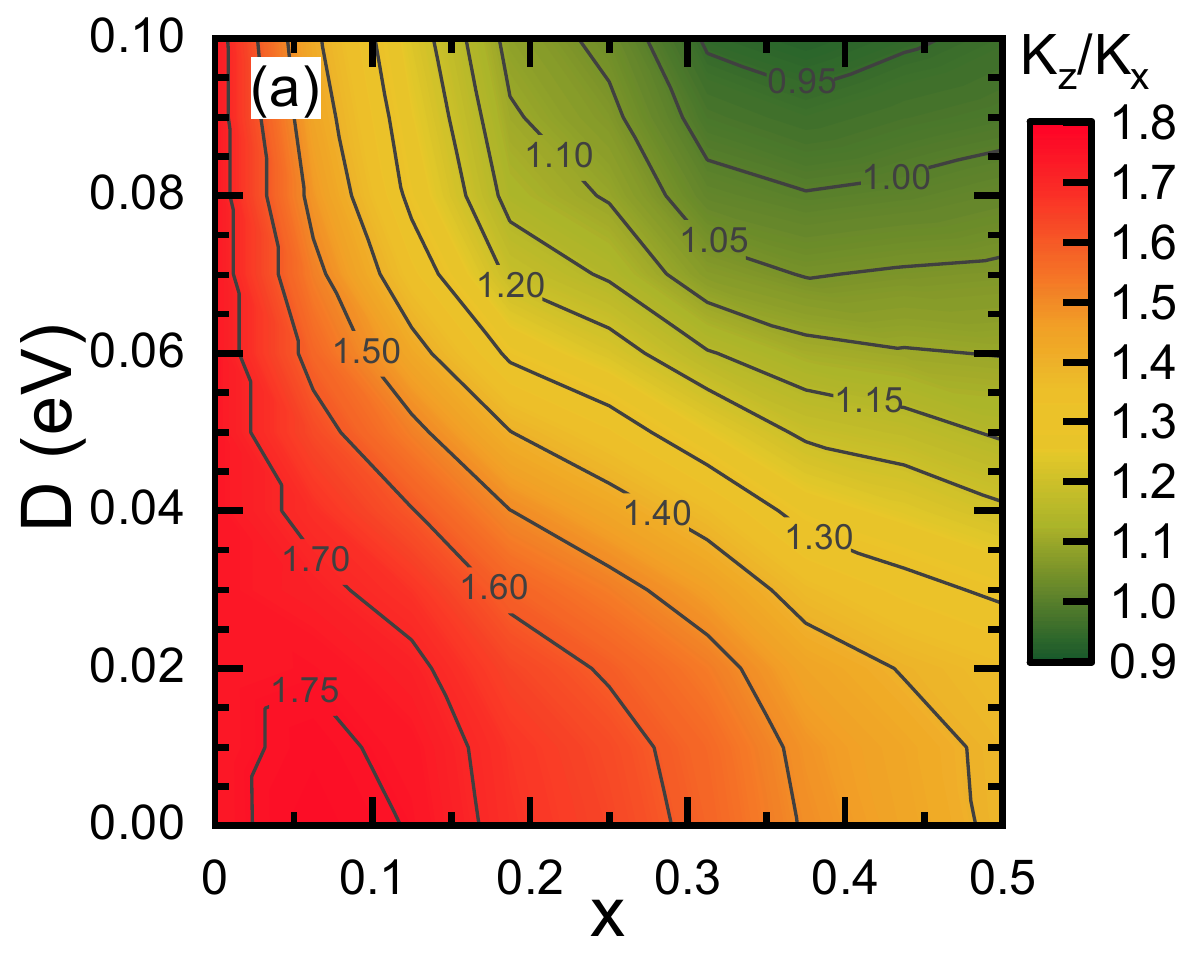}
\includegraphics[width=8cm]{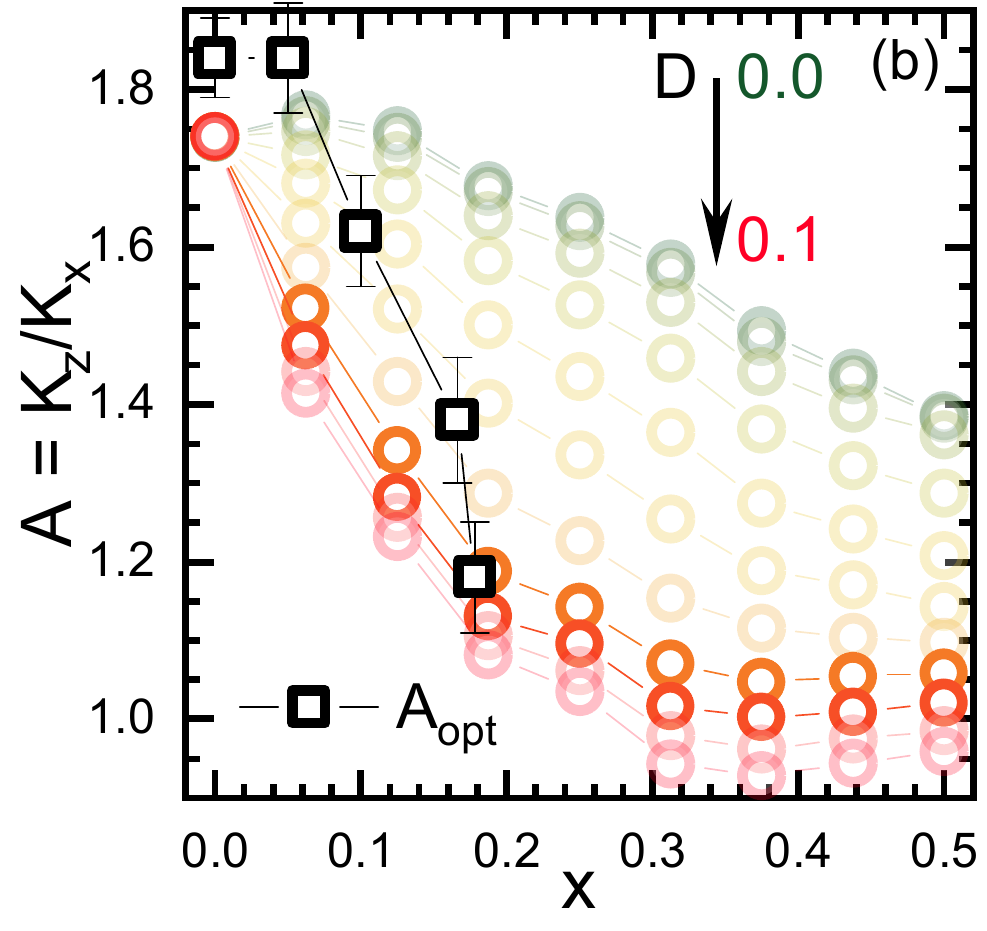}
\caption{Anisotropy of the kinetic energy for increasing doping
\mbox{${\sf x}\in[0,0.5]$:}
(a) Contour plot of the ratio $A\!=\!K_{z}/K_{x}$ for variable defect
induced orbital rotation coupling constant ${\cal D}$, and %
\mbox{%
(b) a comparison%
} of the ${\sf x}$-dependence of the anisotropy $A_{opt}$ obtained
from optical experiments \citep{Fuj06} (squares) for La$_{1-{\sf x}}$Sr$_{\sf x}$VO$_{3}$
and $A$ for %
\mbox{%
${\cal D}=0,0.01,...,0.1$ (legend).%
} Parameter: ${\cal D}_{h}=0$. \label{fig:5}}
\end{figure}

The magnetic anisotropy of the CS/GO phase of the parent compounds
is a manifestation of the strong orbital fluctuations along the $z$-axis
\citep{Kha01,Hor08,Zho09}. This implies a much larger virtual kinetic
energy of $a$ and $b$ orbitals along $z$ as compared to the virtual
hopping along $x$. For the undoped parent state, we find a large
anisotropy, $A\equiv K_{z}/K_{x}\sim1.78$ although the hopping element
$t=0.2$ eV in ${\cal H}_{{\rm Hub}}$ (1) is identical for all cubic
directions. This anisotropy $A$ is similar to the ratio of measured
optical weights $A_{opt}\sim1.84$ for LaVO$_{3}$ \citep{Fuj06}.

Next, we explore the change of the kinetic energy components $K_z$ and
$K_x$ as functions of ${\sf x}$ and ${\cal D}_h$ for fixed
${\cal D}=0.05$ eV. The upward shift of $K_z$ with hole-doping ${\sf x}$
in Fig.~\ref{fig:4}(a) reflects the loss of super-exchange or binding
energy. A loss is further amplified by the ${\cal D}_h$ term. The total
$K_x$ values in Fig.~\ref{fig:4}(b) lie close to $-0.06$ eV and show
only a marginal dependence on ${\sf x}$ and ${\cal D}_h$. The small
changes of the components $K_x^{ab}$ and $K_x^c$ with ${\cal D}_h$
reflect the transfer of holes from $\{a,b\}$ to $c$ orbitals. Hence, we
find that the change of the total kinetic energy $K$ with doping
${\sf x}$ is almost completely determined by $K_{z}$, which results
from large $\{a,b\}$ orbital fluctuations along $z$ that favor $G$-type
orbital order. The data in Fig.~\ref{fig:4} implies a decrease of the
anisotropy $A$ with increasing polaron parameter ${\cal D}_h$
(at ${\cal D}\!=0.05$ eV), however most of the reduction of $A$ with
doping results from ${\cal D}$, i.e., the OR clouds induced by the
defects. It is worth noting that, it is much harder to reach
self-consistency at moderate ${\cal D}_h$---due to the motion of
holes---than in calculations with defect-induced ORs alone.

The contour plot of the anisotropy $A\!=\!K_{z}/K_{x}$ of kinetic
energies in Fig.~\ref{fig:5}(a) for ${\cal D}_{h}\!=\!0$ displays
a strong reduction towards $A\!\sim\!1$ when both the defect induced
OR parameter ${\cal D}$ and doping ${\sf x}$ become sufficiently
large. At $\,{\sf x}\!=\!0.2$ and ${\cal D}\!=\!0.06$ eV, the
asymmetry $A\!\sim\!1.3$, whereas in absence of ORs, for
${\cal D}\!=\!0\!=\!{\cal D}_{h}$, the decay of $A$ with ${\sf x}$ is
much weaker. In the latter case, it is caused exclusively by the motion
of doped holes bound to defects as small spin-orbital polarons
\citep{Ave18}. Figure \ref{fig:5}(b) shows the decay of $A$ versus
doping ${\sf x}$ for different values of ${\cal D}$ and provides a
comparison with the optical anisotropy $A_{opt}$ as determined for
La$_{1-{\sf x}}$Sr$_{\sf x}$VO$_{3}$ \citep{Fuj05}, the only system
where such data seems to exist. It is remarkable that the
theoretical $A$ and experimental $A_{opt}$ almost coincide for the
undoped system. The most pronounced discrepancy is a tendency of
$A_{opt}$ towards a cooperative transition.

However, the cooperative nature of the decay of orbital order in the
(La,Sr) system appears as an exception; experiments for (Pr,Ca),
(Nd,Sr), and (Y,Ca) show a gradual decline of the order parameter with
${\sf x}$ \citep{Fuj05,Ree16}. We saw above that ORs induced by defects
act non-cooperatively, consistent with such a gradual decline. Yet, ORs
induced by polaron charges may well lead to cooperative transitions as
they are driven by $e$-$e$ interactions; accordingly, they can be
extremely relevant to fine tune the theory to specific compounds.

\begin{table}[b!]
\caption{Doping dependence of occupation numbers $n_{ab}$ and $n_{c}$ in
the atomic limit ($t\rightarrow0$) for different limiting cases for
${\cal D}_h$ and ${\cal D}$. Here, ${\cal D}_h^{sat}$ and ${\cal D}^{sat}$
denoted saturation values where the orbital rotation is complete.
\label{tab:para}}
\vskip .1cm \begin{ruledtabular} %
\begin{tabular}{cccc}
${\cal D}_{h}$ & ${\cal D}$ & $n_{ab}$ & $n_{c}$ \tabularnewline
\colrule
$0$  & $0$  & $1-{\sf x}$  & $1$ \tabularnewline
${\cal D}_{h}^{sat}$  & $0$  & $1+{\sf x}$ & $1-2{\sf x}$ \tabularnewline
$0$ & ${\cal D}^{sat}$ & $1-{\sf x}+\frac83{\sf x}$ & $1-\frac83{\sf x}$ \tabularnewline
\end{tabular}\end{ruledtabular}
\end{table}

\section{Conclusions}
\label{sec:conclusions}

The robustness of the insulating state and of the $G$-type orbital
order in the vanadates, observed in several experiments \citep{Fuj05},
has two main causes: \hfill\break
(i) Doped holes are localized by defects and form small
spin-orbital-polarons. The main kinetic energy gain of a doped hole
is a double exchange process on an {\it active bond}
(a ferromagnetic bond in c-direction) next to a defect. \hfill\break
(ii) Orbital order is predominantly suppressed as function of doping
${\sf x}$ not by the kinetic energy associated with the small polaron
but rather by non-cooperative orbital rotations induced by defect
charges.  \hfill\break
An important feature of both orbital
rotation mechanisms, namely the orbital rotations due to the charges
of defects and of polarons, is the transfer of holes from the $a/b$
orbitals to $c$ orbitals. In the absence of these terms, in the atomic
limit $t\rightarrow0$, one finds $n_c=1$ and $n_{ab}=1-{\sf x}$. Where
the latter relation shows that doped holes go into $a/b$ orbitals.
Table I summarizes the transfer of holes in two further limits; for
instance, if ${\cal D}_h$ is in the saturation regime (and ${\cal D}=0$)
the number of electrons in the $a/b$ sector becomes even larger than
one, that is $n_{ab}=1+{\sf x}$, whereas there are now even more holes
in the $c$-orbitals, i.e., $n_c=1-2{\sf x}$, than expected from the
doping concentration ${\sf x}$.

\begin{figure}[t!]
\noindent \centering{}\includegraphics[width=0.97\columnwidth]{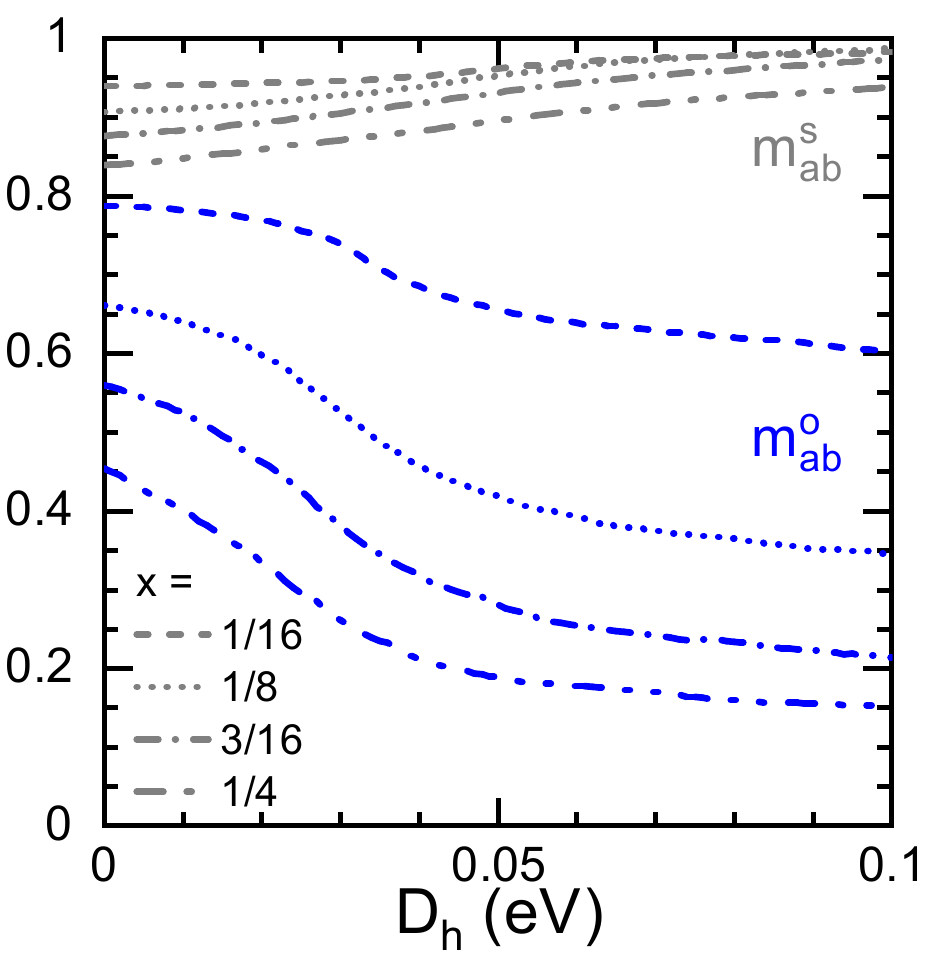}\caption{Decline of orbital order $m_{ab}^{o}$ for several doping concentrations
versus orbital rotation coupling constant $D_{h}$ due to polaron
charges and the simultaneous strengthening of CS order $m_{ab}^{s}$
of $a/b$ electrons, i.e., resulting from the transfer of holes from
$a,b$ to $c$ orbital sector. \label{fig:2SM}}
\end{figure}

Figure~\ref{fig:2SM} displays the decay of the $G$-type orbital order
parameter $m_{ab}^{o}$ as function of ${\cal D}_h$. Interestingly, it
also shows a slight increase of the complementary $C$-type spin order
in the $ab$ orbital sector, $m_{ab}^{s}$. This perhaps surprising
increase of $m^s_{ab}$ is related to the transfer of holes from
$\{a,b\}$ to $c$ orbitals due to ORs. It is reminiscent of the peculiar
robustness of the long-range CS order in the large doping regime where
orbital order is still present, but short ranged \citep{Ree16}.


In summary, we explored the competition between the orbital rotations
induced by the polaron charges and those induced by the defect charges,
\textit{a priori} both being of similar importance. We found that
these rotations are the key mechanisms that control the decay of orbital
order of $t_{2g}$ electrons in doped vanadates, ---much more important
than the string mechanism \citep{Liu92,Lee06,Oga08,Mou07} active
in high-$T_c$ superconductors. When they act together, the qualitative
suppression of orbital order appears to be mainly controlled by the
off-diagonal rotations \eqref{pol} due to defect charges. We found
that the energy scale for rotation of orbitals is primarily determined
by quantum orbital fluctuations rather than by classical crystal fields
and Jahn-Teller potentials. It is very surprising that the
suppression of the anisotropy of kinetic energy under increasing doping
${\sf x}$ occurs in the still-anisotropic $C$-type magnetic state.

\acknowledgments

A.~M.~O. acknowledges Narodowe Centrum Nauki (NCN, Poland)
Project No. 2016/23/B/ST3/00839 and is grateful for the Alexander
von Humboldt Foundation Fellowship \mbox{(Humboldt-Forschungspreis).}
A. A. acknowledges support by MIUR under Project PRIN 2017RKWTMY.

\section*{Appendix: Details of the Hamiltonian}
\label{sec:appendix}

In this Appendix, Subsection A, we summarize the detailed, minimal
multi-orbital  Hubbard model for the description of the spin and
orbital degrees of freedom which give rise to the rich phase diagram
of the $R$VO$_3$ vanadates with perovskite structure. We confine
ourselves to the space spanned by the $t_{2g}$ electrons of vanadium.
An extension of the minimal model, i.e.,  by implementation
of the Coulomb fields of defects and the electron-electron interactions
in Subsection B  allows to study the doped systems.
By the inclusion of the $e$-$e$ interactions the extra screening of
the defect potentials by the doped electrons and/or holes  are
taken into account within the UHF approach.
In Subsection C and D we discuss the derivation of
the extra quadrupolar terms that result from the Coulomb fields of
defects and among electrons, respectively.
Perhaps it is useful here to remind the reader that the monopole Coulomb
fields due to defects and e-e interactions are explicitely contained
in the minimal model as described in Appendix B.

Our general approach follows a similar route as the many-body
treatment of high-$T_c$ superconductors where one also starts from a
minimal model, namely the planar one-band Hubbard model \citep{And87}
---although the model for the vanadates here is more complex.
The Hubbard model of cuprates contains the spin-$\frac{1}{2}$ Heisenberg
model, the doped holes and their interactions. All states not directly
related to Cu$(d^9)$, in particular the O($2p$) states, have been
"integrated out". Their effect is still present in the form of
renormalized effective parameters, like for instance the hopping
parameters $t_{ij}$. As long as one is interested in the low energy and
low temperature physics this approach is fully justified. Only if one
considers spectroscopies in the energy window of $pd$-transitions,
where the  oxygen states come into play, then the model is not
sufficient to describe those, as they have been integrated out.

\subsection{The three-orbital flavor Hubbard model}

The three-orbital Hubbard model for the $t_{2g}$ electrons was
introduced for the triangular lattice \citep{Pen97} and adopted later
for the pnictide superconductors \citep{Dag10}. It has very rich
physics for electronic orders as shown recently \cite{Yue20}.  Here,
we use it for doped vanadium (La,Y)$_{1-{\sf x}}$Ca$_{\sf x}$VO$_{3}$
perovskites \citep{Hor11,Ave13,Ave18},
\begin{equation}
{\cal H}_{{\rm Hub}}={\cal H}_{{\rm kin}}+{\cal H}_{{\rm int}}
+{\cal H}_{{\rm CF}}+{\cal H}_{{\rm JT}}.\label{3band}
\end{equation}
Its main part consists of the kinetic energy ${\cal H}_{{\rm kin}}$
and of the local interactions between the electrons in the three
$t_{2g}$ orbitals, ${\cal H}_{{\rm int}}$. It describes the situation
in the vanadium perovskites after supplementing it by rather weak terms
\citep{Ole07,Ros18}:
the crystal field (CF) splitting ${\cal H}_{{\rm CF}}$, and
the Jahn-Teller (JT) interactions ${\cal H}_{{\rm JT}}$ \citep{Ave13}.

Below we use the definition of $t_{2g}$ orbital degrees of freedom
which selects uniquely a single cubic direction along which
the hopping is inactive to label each orbital flavor \citep{Kha00},
\begin{equation}
|a\rangle\equiv|yz\rangle,\hskip.7cm
|b\rangle\equiv|zx\rangle,\hskip.7cm
|c\rangle\equiv|xy\rangle.
\end{equation}
The kinetic energy for $t_{2g}$ electrons preserves the orbital flavor
in the hopping along the bond $\langle ij\rangle\parallel\gamma$
oriented along one of the cubic axes, $\gamma\in\{a,b,c\}$. It reads,
\begin{equation}
{\cal H}_{{\rm kin}}=\sum_{{\langle ij\rangle\parallel\gamma\atop \alpha\sigma}}t_{ij}^{\gamma\alpha}\left(
\hat{d}_{i\alpha\sigma}^{\dagger}\hat{d}_{j\alpha\sigma}^{}
+\hat{d}_{j\alpha\sigma}^{\dagger}\hat{d}_{i\alpha\sigma}^{}\right).
\label{Ht}
\end{equation}
Here, $\hat{d}_{i\alpha\sigma}^{\dagger}$ is the electron creation
operator in the $t_{2g}$ orbital $\alpha\in\{xy,yz,zx\}$ with spin
$\sigma=\uparrow,\downarrow$ at site $i$. The effective hopping
($t>0$)
\begin{equation}
t_{ij}^{\alpha\beta}=-t\,\delta_{\alpha\beta}
\left(1-\delta_{\gamma\alpha}\right)
\end{equation}
occurs
in two steps, via the hybridization to an intermediate oxygen $2p_{\pi}$
orbital, along idealized 180° V--O--V bonds. Therefore, the hopping is:
(i) diagonal and conserves the orbital flavor $\alpha$ when
the hybridization with the oxygen $2p_{\pi}$
orbitals is finite, and 
(ii) zero otherwise, i.e., $t_{ij}^{\gamma\alpha}=0$ if the
hybridization with the oxygen $2p_{\pi}$ orbitals vanishes by symmetry.

Local interactions at vanadium ions, ${\cal H}_{{\rm int}}$, are
rotationally invariant in the orbital space \citep{Ole83} and depend
on two Kanamori parameters: (i) intraorbital Coulomb interaction $U$
and (ii) Hund's exchange $J_{H}$ between each (equivalent) pair of
$t_{2g}$ electrons in different orbitals,
\begin{align}
{\cal H}_{{\rm int}} & =U\sum_{i\alpha}\hat{n}_{i\alpha\uparrow}\hat{n}_{i\alpha\downarrow}
+J_{H}\sum_{i,\alpha\neq\beta}
\hat{d}_{i\alpha\uparrow}^{\dagger}\hat{d}_{i\alpha\downarrow}^{\dagger}
\hat{d}_{i\beta\downarrow}^{}\hat{d}_{i\beta\uparrow}^{}\nonumber \\
&+\sum_{i,\alpha<\beta}\left[\left(U-\frac{5}{2}J_{H}\right)
\hat{n}_{i\alpha}\hat{n}_{i\beta}
-2J_{H}\hat{\vec{S}}_{i\alpha}\!\cdot\!\hat{\vec{S}}_{i\beta}\right]\!.
\end{align}
Interorbital Coulomb interactions $\propto\hat{n}_{i\alpha}\hat{n}_{i\beta}$
are expressed in terms of orbital electron density operators for a
pair %
\mbox{%
$\alpha<\beta$%
}, %
\mbox{%
$\hat{n}_{i\alpha}^ {}=\sum_{\sigma}\hat{n}_{i\alpha\sigma}^{}=
\sum_{\sigma}\hat{d}_{i\alpha\sigma}^{\dagger}\hat{d}_{i\alpha\sigma}^ {}$%
}. Orbital spin operators, %
\mbox{%
$\hat{\vec{S}}_{i\alpha}\equiv
\{\hat{S}_{i\alpha}^{x},\hat{S}_{i\alpha}^{y},\hat{S}_{i\alpha}^{z}\}$%
}, appear in the Hund's exchange term, %
\mbox{%
$-2J_{H}\hat{\vec{S}}_{i\alpha}\!\cdot\!\hat{\vec{S}}_{i\beta}$.%
} In a Mott insulator, charge fluctuations are quenched and electrons
localize due to large energy of the fundamental Mott gap,
$(U-3J_{H})\gg t$, associated with high-spin charge excitation. In the
case of LaVO$_{3}$, one finds the ground state in a $t_{2g}^2$
configuration at each vanadium ion. Hund's exchange $J_H$ stabilizes
high-spin states with spin $S=1$. The insulating ground state of
LaVO$_3$ has a $C$-type antiferromagnetic spin (CS) order coexisting
with $G$-type alternating orbital (GO), i.e., CS/GO order \citep{Ole07}.

The CF Hamiltonian,
\begin{equation}
{\cal H}_{{\rm CF}}=-\Delta_{c}\sum_{i}\hat{n}_{ic},\label{Hcf}
\end{equation}
lifts the degeneracy of the three $t_{2g}$ orbitals, breaks the cubic
symmetry in the orbital space, and favors the electron occupancy in
$c\equiv xy$ orbitals. This symmetry breaking occurs at the structural
transition intervening at temperature $T_{s}$ \citep{Miy06}. We
take $\Delta_{c}$ as a constant parameter independent of temperature;
it selects the orbital doublet as orbital degree of freedom and gives
either $c_{i}^{1}a_{i}^{1}$ or $c_{i}^{1}b_{i}^{1}$ configuration
at the V ion at site $i$, depending on the actual lattice distortion
in the $ab$ plane. In a Mott insulator, spin-orbital superexchange
explains the ground state observed in LaVO$_{3}$ \citep{Kha01}.

Lattice distortions change the electronic state and induce weak JT
interactions in the three-band Hubbard model (\ref{3band}),
\begin{align}
{\cal H}_{{\rm JT}} & =\frac{1}{4}\,V_{ab}\sum_{\langle ij\rangle{\parallel}ab}(\hat{n}_{ia}-\hat{n}_{ib})(\hat{n}_{ja}-\hat{n}_{jb})\nonumber \\
 & -\frac{1}{4}\,V_{c}\sum_{\langle ij\rangle{\parallel}c}(\hat{n}_{ia}-\hat{n}_{ib})(\hat{n}_{ja}-\hat{n}_{jb}).
\end{align}
Using the orbital $\tau_{i}^{z}$ operators,
\begin{equation}
\tau_{i}^{z}\equiv\frac{1}{2}\sum_{\sigma}\left(\hat{d}_{ia\sigma}^{\dagger}\hat{d}_{ia\sigma}^ {}-\hat{d}_{ib\sigma}^{\dagger}\hat{d}_{ib\sigma}^{}\right),\label{tauz}
\end{equation}
the JT interactions in Eq. (\ref{3band}) are,
\begin{equation}
{\cal H}_{{\rm JT}}=V_{ab}\sum_{\langle ij\rangle{\parallel}ab}\hat{\tau}_{i}^{z}\hat{\tau}_{j}^{z}-V_{c}\sum_{\langle ij\rangle{\parallel}c}\hat{\tau}_{i}^{z}\hat{\tau}_{j}^{z}.
\end{equation}
These interactions stabilize another competing type of spin-orbital
order \citep{Kha01}, the $G$-type AF spin (GS) spin coexisting with
$C$-type AO (CO) order, which represents the GS/CO ground state in
YVO$_{3}$ \citep{Hor11,Fuj10}. Small doping $x\simeq0.01$ leads
to a phase transition to the CS/GO phase \citep{Hor11},
which is the phase studied in this work.

Following the earlier studies \citep{Ave18}, we have fixed the small
parameters in ${\cal H}_{{\rm CF}}$ and ${\cal H}_{{\rm JT}}$ as
follows: $\Delta_{c}=0.1$, \mbox{$V_{ab}=0.03$,} and $V_{c}=0.05$ (all
in eV). The term $\propto V_{ab}$ favors alternating $\{a,b\}$ orbitals,
i.e., $G$-AO order in the $ab$ planes ($V_{ab}>0$) while the
\emph{ferro-orbital} order is favored along the $c$ cubic axis
($V_{c}>0$). Thus, the term $\propto V_{c}$ weakens the superexchange
orbital interaction $\propto Jr_{1}$, where $J=4t^{2}/U$ and
$r_1=(1-3\eta)^{-1}$ with $\eta=J_{H}/U$ \citep{Kha01}. One finds that
for the present parameters ($U=4.5$, $t=0.2$, $J_{H}=0.5$, all in eV)
$Jr_{1}=53$ meV, so taking $V_{c}=50$ meV one is indeed close to the
switching of the orbital order observed in YVO$_{3}$
\citep{Fuj10,Sah17,Yan19}.

\subsection{Coulomb fields and orbital rotations due to charged defects and polarons}

The complete Hamiltonian for $t_{2g}$ electrons in doped vanadium
(La,Y)$_{1-{\sf x}}$Ca$_{\sf x}$VO$_{3}$ perovskites, Eq.~(1), reads
\citep{Ave18},
\begin{equation}
{\cal H}_{t2g}\!={\cal H}_{{\rm Hub}}+\sum_{i<j}v(r_{ij})\hat{n}_{i}\hat{n}_{j}
+\sum_{mi}v(r_{mi})\hat{n}_{i}+{\cal H}_{{\rm pol}}.\label{Ht2g}
\end{equation}
It includes the three-band Hubbard model ${\cal H}_{{\rm Hub}}$ \citep{Ole07}--
see above --, the long-range electron-electron and defect-electron
interactions $\propto v(r)$ -- see main text -- and orbital polarization
terms,
\begin{equation}
{\cal H}_{{\rm pol}}={\cal H}_{{\rm pol}}^{(1)}+{\cal H}_{{\rm pol}}^{(2)}.
\end{equation}

The first term was analyzed before and is responsible for the collapse
of orbital order under doping by charged defects~\citep{Ave19}.

\subsection{Defect induced orbital polarization}

The term of the Hamiltonian for vanadium ions at $\mathbf{r}_{i}$
in the Coulomb potential of defects $D^{-}$ at $\mathbf{R}_{m}$
is:
\begin{equation}
{\cal H}_{\cal D}\!=\sum_{{m,i\in\mathcal{C}_{m}\atop \alpha,\beta,\sigma}}\!\!\langle i\alpha|v(|\mathbf{r}\!-\!\mathbf{R}_{m}|)|i\beta\rangle\hat{d}_{i\alpha\sigma}^{\dag}\hat{d}_{i\beta\sigma}^ {}.\label{HD}
\end{equation}
Introducing
\begin{equation}
{\cal V}_{mi}^{\alpha\beta}=\langle i\alpha|v(|\mathbf{r}\!-\!\mathbf{R}_{m}|)|i\beta\rangle
\end{equation}
and
\begin{equation}
{\bar{{\cal V}}}_{mi}^{\alpha\alpha}=\frac{1}{3}\sum_{\alpha}{\cal V}_{mi}^{\alpha\alpha}\equiv v(r_{mi})
\end{equation}
one obtains
\begin{equation}
{\cal H}_{\cal D}\!=\sum_{mi}v(r_{mi})\hat{n}_{i}+{\cal H}_{{\rm pol}}^{(1)}.\label{HD2}
\end{equation}
where the sums in the first term extend over the whole system.

The range of the polarization term, due to the short-range nature of the
matrix elements, will be restricted to the defect cube of the respective
defect $\mathcal{C}_{m}$. The polarization term
\begin{equation}
{\cal H}_{{\rm pol}}^{(1)}\!=\sum_{{mi\in\mathcal{C}_{m}\atop \alpha\sigma}}
\left({\cal V}_{mi}^{\alpha\alpha}-v(r_{mi})\right)\hat{n}_{i\alpha\sigma}
+\!\sum_{{mi\in\mathcal{C}_{m}\atop \alpha\neq\beta,\sigma}}\!{\cal V}_{mi}^{\alpha\beta}\hat{d}_{i\alpha\sigma}^{\dag}\hat{d}_{i\beta\sigma}^{}.
\label{H1pol}
\end{equation}
consists of a diagonal and an off-diagonal terms. For a non-distorted
cubic neighborhood only the off-diagonal terms contribute. where the
coupling constant ${\cal D}$ is defined by the matrix element
\begin{equation}
{\cal D}\lambda_{\alpha\beta}^{{\bf d}}\equiv{\cal V}_{mi}^{\alpha\beta}=\langle i\alpha|v(|\mathbf{r}\!-\!\mathbf{R}_{m}|)|i\beta\rangle.\label{Voff}
\end{equation}
Here $\lambda_{\alpha\beta}^{{\bf d}}$ contains the signs of the
matrix elements and is displayed in the main text. The signs do depend
on the respective diagonal of the defect cube, i.e., parallel to the
vector $\mathbf{d}\!=\!\mathbf{r}_{i}\!-\!\mathbf{R}_{m}$ connecting
the respective V-ion and the defect. The rotation operator can then
be summarized as
\begin{equation}
\mathcal{H}_{{\rm pol}}^{(1)}={\cal D}\!\sum_{{m,i\in\mathcal{C}_{m}\atop \alpha,\beta,\sigma}}\lambda_{\alpha\beta}^{\mathbf{d}}\;
\delta_{\mathbf{d},\mathbf{r}_{i}\!-\!\mathbf{R}_{m}}
\hat{d}_{i\alpha\sigma}^{\dag}\hat{d}_{i\beta\sigma}^{}.\label{pol}
\end{equation}
A value ${\cal D}\approx50$ meV has been estimated by simple
defect-potential-mediated
superposition integrals, given a strength of the defect potential
at the vanadium sites (at a given $t_{2g}$ orbital) of 2 eV.

\subsection{Orbital polarization due to the polaron charge}

The orbital polarization induced on the vanadium ion at $\mathbf{r}_{i}$
due to the charge of polaron at $\mathbf{r}_{j}$ stems from e-e interactions.
The leading term ${\cal H}_{{\rm pol}}^{(2)}$ has monopole-quadrupole
character. Similar to the defect case we can write
\begin{equation}
{\cal H}_{{\rm pol}}^{(2)}\!=\sum_{{ji\in\mathcal{C}_{j}\atop \alpha,\beta,\sigma}}
\left({\cal V}_{ji}^{\alpha\beta}-\delta_{\alpha\beta}v(r_{ji})\right)
\left(\langle n_{j}\rangle-n_{0}\right)
\hat{d}_{i\alpha\sigma}^{\dag}\hat{d}_{i\beta\sigma}^{}.\label{H2pol}
\end{equation}
Here $(\langle n_{j}\rangle-n_{0})$ represents the negative density
of doped holes, and $n_{0}$ is the number of electrons per V-ion
in the undoped case, i.e., for the $d^{2}$ configuration $n_{0}=2$.
The coupling constant ${\cal D}_h$ follows from the two-center matrix
element
\begin{equation}
{\cal V}_{ji}^{\alpha\beta}=\langle i\alpha|v(|\mathbf{r}\!-\!\mathbf{r}_{j}|)|i\beta\rangle.\label{Vdiag}
\end{equation}
For the undistorted cubic system only diagonal terms contribute, that
depend on the vector $\mathbf{h}\!=\!\mathbf{r}_{i}\!-\!\mathbf{r}_{j}$
connecting V-ion and the polaron density at $\mathbf{r}_{j}$, where
we include only the nearest V-ions in $\mathcal{C}_{j}$ . The
coupling constant ${\cal D}_h$ is defined by the matrix element of the
field of the hole at $\mathbf{r}_{j}$ and the orbitals at $\mathbf{r}_{i}$:
\begin{equation}
{\cal D}_{h}\zeta_{\alpha\beta}^{{\bf h}}\equiv\left(\langle i\alpha|v(|\mathbf{r}\!-\!\mathbf{r}_{j}|)|i\beta\rangle-v(|\mathbf{h}|)\right)
\delta_{\alpha\beta}.
\end{equation}
We find that for an undistorted cubic lattice $\zeta_{\alpha\beta}^{{\bf h}}$
is diagonal with respect to the global $t_{2g}$ basis
\begin{equation}
\zeta_{\alpha\beta}^{{\bf h}}=\delta_{\alpha\beta}\left(3\delta_{\alpha\gamma}-1\right)
\label{zeta}
\end{equation}
and depends on the cubic axis $\gamma({\bf h)}$, i.e., parallel to
the vector $\mathbf{h}\!=\!\mathbf{r}_{i}\!-\!\mathbf{r}_{j}$ that
connects the polaron density at $\mathbf{r}_{j}$ and a n.n. V-ion
at $\mathbf{r}_i$. In our study, we shall use either ${\cal D}=0$ or
50 meV with ${\cal D}_h$ as a free parameter. Alternatively, we use the
geometric relation ${\cal D}_h/{\cal D}\simeq(d/a)^{\gamma}$ defined by
the nearest neighbor V-V and the V-D distance, labeled as $a$ and $d$,
\mbox{respectively.} This corresponds to 0.87 for $\gamma=1$, which we
use here.




%

\end{document}